\documentclass{PoS}

\PoS{PoS(LAT2005)094}
\usepackage{graphicx}

\newcommand{\bea}{\begin{eqnarray}}
\newcommand{\eea}{\end{eqnarray}}
\newcommand{\be}{\begin{equation}}
\newcommand{\ee}{\end{equation}}

\newcommand{\mlog}{{\rm log}}

\newcommand{\ce}{\chi_\eta}
\newcommand{\ci}{\chi_{l}}

\newcommand{\ciii}{\chi_{s}}

\title{Probing the chiral limit of $M_\pi$ and $f_\pi$ in 2+1 flavor QCD with
  domain wall fermions from QCDOC}

\ShortTitle{Probing the chiral limit of $M_\pi$ and $f_\pi$ in 2+1 flavor QCD with DWF }

\author{\speaker{Meifeng Lin} \\
        Department of Physics\\ Columbia University\\ New York, NY 10027\\ USA\\
        E-mail: \email{mflin@physics.columbia.edu}}

\author{RBC and UKQCD Collaborations}

\abstract{We present results for the pseudoscalar meson masses and decay constants 
on 2+1 flavor DWF configurations with different sea quark masses and 
an inverse lattice spacing of 1.6(1) GeV, with a focus on chiral 
fits at small quark masses. The calculation is done on $16^3\times32\times8$ 
lattices generated with the DBW2 gauge action.}

\FullConference{XXIIIrd International Symposium on Lattice Field Theory\\
                 25-30 July 2005\\
                 Trinity College, Dublin, Ireland}

\begin{document}

\section{Introduction}
Lattice calculations allow us to study hadronic properties from first
principles. One of the great challenges of lattice QCD, however, is that the
physical light quark masses, in particular the physical up and down sea quark masses, are
numerically inaccessible with present hardware. We thus rely on (partially
quenched) chiral extrapolations to get to the physical light quark regime
from simulations with unphysically heavy quarks. We can also determine the low
energy constants (LECs)~\cite{GL85}, \textit{aka} Gasser-Leutwyler coefficients,
which appear in the chiral
Lagrangian, from these partially quenched simulations provided the systematic
errors are treated properly and the correct number of fermion flavors are used~\cite{Sharpe00}. 

Domain wall fermions (DWF) have the advantages of nearly
exact chiral symmetries with a moderate extent in the fifth dimension and a
continuum-like chiral perturbation theory. The residual chiral symmetry
breaking of DWF, quantified as $m_{\rm res}$~\cite{Spectrum00}, comes in as an additive constant
to the quark masses at the lowest order approximation. The employment of the rational hybrid Monte Carlo (RHMC)
technique~\cite{RHMC05}, free of finite step-size errors, also gives us better control over systematic errors. It thus makes possible a
direct comparison of our numerical results with the predictions of chiral
perturbation theory. 
This proceeding focuses on a set of $N_f = 2+1$ DWF simulations on
$16^3\times32\times8$ lattices with a lattice spacing of 1.6(1) GeV. After detailing the simulation parameters, I present
results for the masses and decay constants of the pseudoscalar
mesons. Preliminary chiral fits up to next-to-leading order (NLO) on these
quantities are then given. A rough investigation of the effect of
nondegeneracy is also described. 
\vspace{-2mm}
\section{Simulation details}
The gauge action used in our simulation can be written in the
general form~\cite{DBW2}: 
\begin{equation}
  S_G[U] = -{{\beta}\over{3}}\left(\left(1-8c_1\right)\sum_{x;\mu,\nu}P[U]_{x,\mu\nu}+c_1\sum_{x;\mu\neq\nu}R[U]_{x,\mu\nu}\right)
\label{eqn:gauge}
\end{equation}
We generated three sets of $16^3\times32\times8$ gauge configurations with the DBW2
action, in which $c_1 = -1.4069$, at $\beta = 0.72$.
 Two light sea quarks with equal mass $m_{l}$ and one strange quark with mass
 $m_s$ were included in
the fermion determinant. The RHMC algorithm was used to update the gauge
fields. All the data was generated on QCDOC machines at Columbia, the RBRC and
Edinburgh.

Valence measurements with two degenerate quarks of up to 8 different masses $m_V$ were
performed on these lattices. Nondegenerate valence measurements with light
quark masses of 0.005, 0.01, 0.015, 0.02 and a heavy quark mass of 0.04 were
also done for $m_{l} = 0.01$ and $0.02$. We skipped the first 1000 trajectories for
thermalization and measured \emph{point-point} meson correlators thereafter. Table~\ref{tab:param} shows the relevant parameters
for evolution and measurements. The integrated autocorrelation time was determined to be on the
order of 50 to 100 trajectories. Thus we binned the data into blocks of 100
trajectories prior to statistical analysis to account
for the autocorrelations in a robust fashion. All the quantities are in
lattice units unless noted.
\begin{table}[t]
\centering
\caption{Simulation parameters. $\delta\tau$ is the step size in molecular
dynamics integration and $\tau$ is the total length in one trajectory. Data for the $m_{l} = 0.01$ and $0.02$ ensembles used in this proceeding reflects twice as much statistics as that in the talk given at the conference.}
\begin{tabular*}{\textwidth}{lllllllll}
\hline
\hline
 ($m_{l}$,$m_s$) & $L_s$ & $M_5$ &$\delta\tau$ & $\tau$ & $\frac {N_{acc}}{N_{tot}}$ & \# traj.
 & $m_{V}$ & \# meas.(deg;nondeg)\\
\hline
 (0.01,0.04) & 8 & 1.8 & 0.02   & 0.5 & 66\% & 6000 & 0.005 $\to$ 0.04 & 500; 250\\
 (0.02,0.04) & 8 & 1.8 & 0.0185 & 0.5 & 70\% & 6000 & 0.005 $\to$ 0.04 & 1000;
 500\\
 (0.04,0.04) & 8 & 1.8 & 0.02   & 0.5 & 65\% & 3395 & 0.02, 0.03, 0.04  & 479;
 --\\
\hline
\end{tabular*}
\label{tab:param}
\end{table}

The residual
chiral symmetry breaking turns out to be quite large at this coupling. When
extrapolated to the dynamical chiral limit ($m_l = 0$), 
\be
m_{\rm res} = 0.0106(1)
\ee
which is comparable to the input light sea quark mass. Noting that
the coupling is fairly large and $L_s$ is rather small, it is not surprising to
have a large $m_{\rm res}$. For hadronic observables like meson masses
and decay constants, we treat $m_{\rm res}$ as a shift to the input quark
masses and neglect other possible higher-order effects.
\section{Preliminary Results}
The pseudoscalar masses and decay constants can be extracted from both the
pseudoscalar and axialvector correlators~\cite{Spectrum00}. Here emphasis will
be given to the results from the pseudoscalar channel, since they give
smaller statistical errors than the axialvector correlators.

\underline{\bf $M_\pi$ and the chiral fits.}
The next-to-leading order (NLO) quark mass dependence of the pseudoscalar
masses in PQ$\chi$PT with $N$ flavors of sea quark 
has been computed generally~\cite{Sharpe00}. With 2+1 flavors of sea
quark and two degenerate valence quarks, the formula simplifies to
 \bea
{M_{\pi}^2}&=& {\chi_V \Big \lbrace 1+ \frac{16N}{f^2}(2 L_6-L_4) \bar\chi
+\frac{16}{f^2}(2L_8-L_5) \chi_V }\nonumber \\
& &{+\frac{1}{8f^2\pi^2 N} \big \lbrack
  \frac{2\chi_V-\ci-\ciii}{\chi_V-\ce}\chi_V\mlog\chi_V } { - \frac{(\chi_V-\ci)(\chi_V-\ciii)}{(\chi_V-\ce)^2}\chi_V\mlog\chi_V}
\nonumber \\
& &{ +\frac{(\chi_V-\ci)(\chi_V-\ciii)}{\chi_V-\ce}(1+\mlog\chi_V)} {+\frac{(\ce-\ci)(\ce-\ciii)}{(\chi_V-\ce)^2}\ce\mlog\ce
\big \rbrack \Big \rbrace }
\label{eq:mass_NLO}
\eea
where $\chi_x = 2 B m_x,\ x = V,\ l,\ s,\ {\rm or} \ \eta,\ 
\ce = \frac{1}{3}(\ci+2\ciii)$ and $\bar{\chi} =
\frac{1}{3}(2\ci+\ciii)$. 
Here we have explicitly taken the chiral scale to be 1 GeV. For DWF, the masses should all be
shifted by $m_{\rm res}$.  Thus to leading order in the valence quark masses,
$M_{\pi}^2 = 2 B (m_V + m_{\rm res})$. 

\begin{figure}[t]
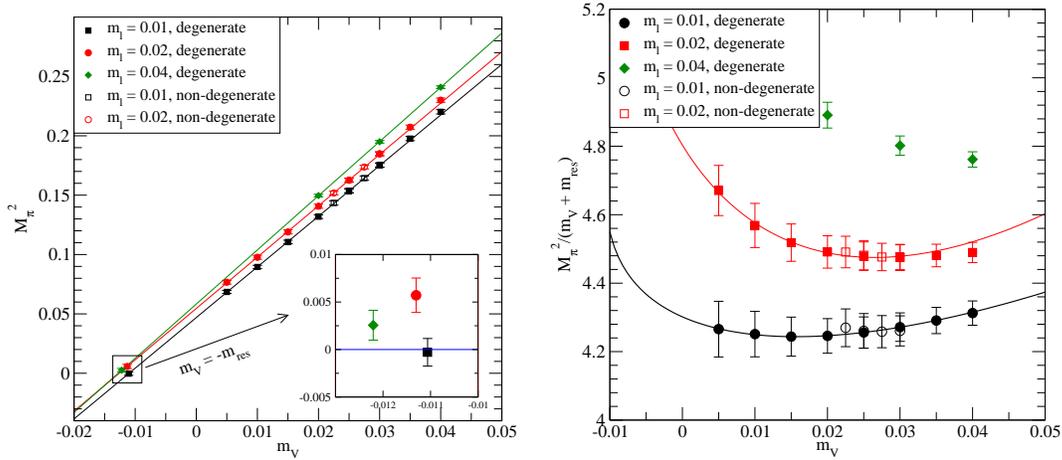

\begin{center}
\begin{tabular}{ll}
\includegraphics*[width=0.45\textwidth,clip]{mpi_linear.eps}&
\includegraphics*[width=0.45\textwidth,clip]{mpi_res.eps}
\end{tabular}
\caption{\textbf{LEFT:} $M_\pi^2$ vs. $m_V$ in the pseudoscalar channel;
  \textbf{RIGHT:} $M_\pi^2/(m_V+m_{\rm res})$ vs. $m_V$. The lines are fits
  to the mass range [0.005,0.03]. Filled symbols and open symbols represent data from
  measurements with degenerate and nondegenerate quarks, respectively. $m_V$ is the
  average of two valence quark masses. }
\label{fig:all}
\end{center}
\end{figure}

As a consistency check, we fit our results for $M_{\pi}^2$ to the linear form
\be
M_{\pi}^2 = 2B(m_V + m_{\rm res}) + C
\label{eq:mass_LO_2}
\ee
where $m_{\rm res}$ is the residual mass at the valence chiral limit ($m_V \to
0$), \textit{i.e.}, 0.0111(1), 0.0113(1) and 0.0122(1) for $m_l=0.01$, $0.02$ and $0.04$
respectively. We would expect $C$ to vanish if the lowest-order approximation is good enough.  Independent linear fits for the pion masses with
different sea quarks are shown in the left panel of Figure~\ref{fig:all}. The two
heaviest masses were excluded from the fit except for $m_l=0.04$ where all the
available data were used. In all the cases, $M_\pi^2$ is
quite close to zero when $m_V = -m_{\rm res}$. The larger deviation from 0 for
$m_l = 0.02$ may be due to higher order corrections from $\chi$PT or ${\cal
  O}(a^2)$ violations of chiral symmetry, which need to be further
investigated.

We also show $M_\pi^2 / (m_V + m_{\rm res})$ as a function of $m_V$ in the right panel of Figure~\ref{fig:all}. Nonlinearities are evident in all
the datasets. The solid lines are the partially quenched NLO fit to
Eq.~\ref{eq:mass_NLO} with sea quark mass fixed. We chose the fitting range to
be [0.005,0.03]. In the fit, the value of $f_\pi$ at
$m_l = m_V = -m_{\rm res}$ from the dynamical linear extrapolation was used as the
input for $f$, while in principle the limit of $m_s\to -m_{\rm res}$ should also be taken
to obtain the value for $f$. As can be seen, for $m_l = 0.01$ the curve does go
through all the data. For $m_l = 0.02$ the NLO fit represents the data well, for  all but
       the heaviest valence masses, where differences of about one
       standard deviation are seen. As will be shown later, the NLO fit to
$f_\pi$ also has the same problem. 

\underline{\bf $f_\pi$ and the chiral fits.}
We use the definition of $f_\pi$ in which its physical value is about
130 MeV. When two degenerate valence quarks are used, the NLO mass dependence of $f_\pi$ can be
simplified to \cite{Sharpe00,Bijnens05}
\bea
f_{\pi}& = &f \Big \lbrace 1 + \frac{8}{f^2} (N L_4 \bar{\chi} + L_5 \chi_V)
- \frac{1}{16 \pi^2 f^2} \Big\lbrack {(\chi_V + \ci)}\mlog\frac{\chi_V+\ci}{2} +
\frac{\chi_V + \ciii}{2} \mlog\frac{\chi_V+\ciii}{2}\Big\rbrack \Big \rbrace
\label{eq:fpi}
\eea

\begin{figure}[t]
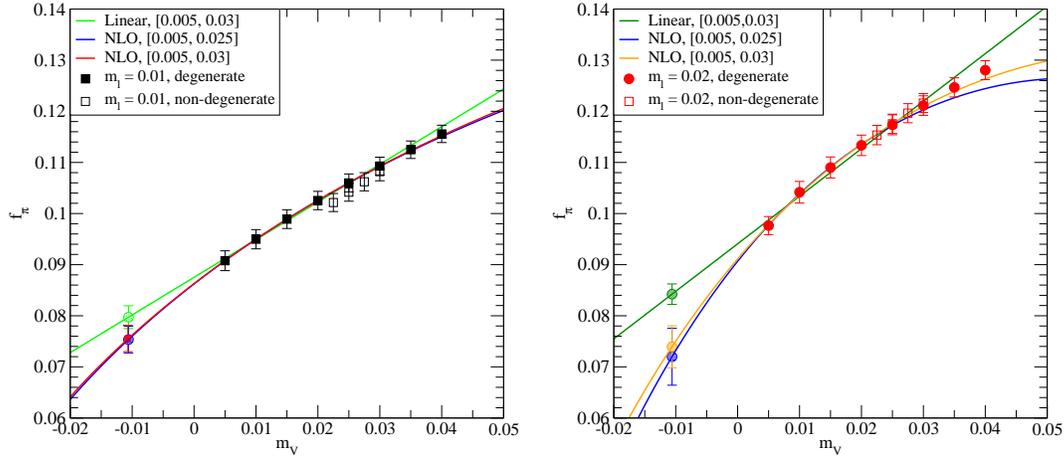

\begin{center}
\begin{tabular}{ll}
\includegraphics*[width=0.45\textwidth,clip]{fpi_mu01.eps} &
\includegraphics*[width=0.45\textwidth,clip]{fpi_mu02.eps}
\end{tabular}
\caption{$f_\pi$ with linear and NLO fits to mass range
  [0.005,0.025] and [0.005,0.03]  for $m_l = 0.01$ (left) and $m_l = 0.02$ (right) datasets. Filled symbols and open symbols represent data from
  measurements with degenerate and nondegenerate quarks, respectively. $m_V$ is the
  average of two valence quark masses.}
\label{fig:fpi}
\end{center}
\end{figure}

We show the results for $f_\pi$ from the pseudoscalar correlators in
Figure~\ref{fig:fpi}. The data points have obvious curvatures, especially for
$m_l = 0.02$. The straight lines fit the data poorly. We also fit the $m_l = 0.01$ and $m_l = 0.02$ data independently to
Eq.~\ref{eq:fpi} using the values of $B$ from the NLO fit of $M_\pi^2$ as an
input. Two fitting ranges were used: [0.005,0.025] and [0.005,0.03]. For $m_l
= 0.01$, the fitting results do not depend on the fitting range we
choose. Even a fit to the first five points reproduces the data at heavier
quark masses quite well. However, this is not the case for $m_l = 0.02$, where
the fit curve misses the data points at heavier masses. It is especially
obvious for the fit to the range [0.005,0.025]. Similar behaviors were
observed in $N_f = 2$ DWF simulations~\cite{2fDWF}, where, instead of fitting
data with different sea quarks independently, simultaneous fits to
$f_\pi$ from different sea quarks were performed. 

\underline{\bf Remarks.}
We cannot conclude here that we have found a signal of chiral logs. The
nonlinearities we see for both $M_\pi^2$ and $f_\pi$ may also be attributed to
finite-volume corrections, zero-mode effects or uncertainties in the residual
mass. For the large quark masses we used, NNLO contributions may already be
important, especially for $f_\pi$. More thorough investigations of these
systematic effects are needed to justify our NLO fits. 

\section{ Effect of nondegeneracy}
\begin{figure}[h]
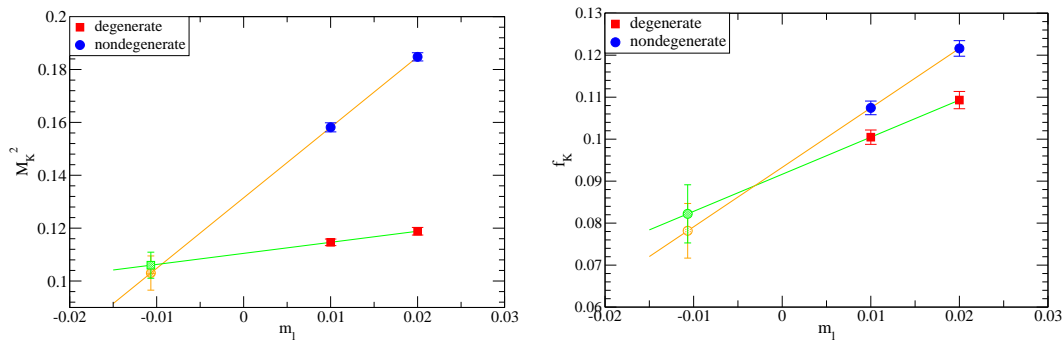

\begin{center}
\begin{tabular}{ll}
\includegraphics*[width=0.45\textwidth,clip]{mK.eps} &
\includegraphics*[width=0.45\textwidth,clip]{fK.eps}
\end{tabular}
\caption{Estimate the effects of nondegeneracy.}
\label{fig:nondeg}
\end{center}
\end{figure}

We also explored the possible effects of nondegenerate quarks in the
pseudoscalar masses and decay constants. Suppose we have two sets of
valence measurements on the same lattices. One uses two degenerate quarks ($m_A = m_B
= m_V$) and the other uses two nondegenerate quarks ($m_A \neq m_B$, $m_V =
\frac{m_A + m_B}{2}$). Leading order $\chi$PT predicts, for the
pseudoscalar masses, $M_\pi^2 = B (m_A + m_B)$. If we take the nondegenerate dynamical points ($m_A = m_{l}$, $m_B = m_s$)
and the matching degenerate points ($m_V + m_{\rm res} = \frac{m_s + m_{\rm res}}  {2}$) to
extrapolate to the light sea quark limit ($m_{l} \to -m_{\rm res}$), we would expect the results
to be the same assuming the higher order corrections are small. This gives us
rough estimates of the kaon mass from both the
degenerate and nondegenerate valence quarks. The same comparison can also be done
for the pseudoscalar decay constants. We show the comparisons in
Figure~\ref{fig:nondeg}. The two extrapolations gave quite consistent
results within errors.

\section{Summary and Outlook}
Our present data are still statistically limited. Further improvements are
needed to reduce the statistical uncertainties in the determinations of $M_\pi$ and
$f_\pi$. In the meanwhile, better estimates of the systematic effects coming
from finite volume, finite lattice spacing and residual chiral symmetry
breaking are necessary to justify our chiral fits and determine LECs. For the work
reported here, the smallest $M_\pi L$ is 4.2, which should make the finite
volume corrections few percent effects. The ongoing $24^3\times64\times16$
simulations with 2+1 dynamical flavors of DWF on QCDOC will provide us better
opportunities to probe the chiral limit of various physical quantities.

\section*{Acknowledgments}
We thank Peter Boyle, Mike Clark, Chris Maynard, Robert Tweedie, and Azusa Yamaguchi
for help generating the datasets used in this work. We thank Peter Boyle, Dong Chen, Norman Christ, Mike Clark,
Saul Cohen, Calin Cristian, Zhihua Dong, Alan Gara, Andrew Jackson, Balint Joo,
Chulwoo Jung, Richard Kenway, Changhoan Kim, Ludmila Levkova, Xiaodong Liao,
Guofeng Liu, Robert Mawhinney, Shigemi Ohta, Konstantin Petrov, Tilo Wettig and Azusa Yamaguchi
for developing with us the QCDOC machine and its software. This development
and the resulting computer equipment used in this calculation were funded
by the U.S. DOE grant DE-FG02-92ER40699, PPARC JIF grant PPA/J/S/1998/00756
and by RIKEN. This work was supported by DOE grant DE-FG02-92ER40699 and
we thank RIKEN, BNL and the U.S. DOE for providing the facilities essential
for the completion of this work.


\end{document}